\documentstyle[aps,prb,epsfig]{revtex}
\begin{document}
\baselineskip=0.5cm
\renewcommand{\thefigure}{\arabic{figure}}
\title{Hartree-Fock-Bogoliubov theory of a charged Bose gas
at finite temperature}
\author{B. Davoudi$^{1,2}$, A. Minguzzi$^1$ and M. P. Tosi$^1$}
\address{
$^1$NEST-INFM and Classe di Scienze, Scuola Normale Superiore, I-56126 Pisa, Italy\\ 
$^2$Institute for Studies in Theoretical Physics and Mathematics, Tehran, P.O.Box 19395-5531,Iran\\
}
\date{\today}
\maketitle
\begin{abstract}
We critically examine the Hartree-Fock-Bogoliubov (HFB) solution of the equations of motion for condensate fluctuations in a weakly coupled plasma of charged bosons at finite temperature. Analytic expressions are derived for the first two infrared-divergent terms in both the momentum distribution of the noncondensate and the anomalous Bose correlation function at low momenta. Incorporation into the theory of the appropriate form of the Hugenholtz-Pines relation for the chemical potential is needed to cancel an unphysical divergence. Exact cancellation of infrared-divergent terms is demonstrated in the HFB shift of the single-particle excitation energy away from the Bogoliubov value at long wavelengths, with the residual terms raising it towards the plasma frequency at low temperature. Numerical illustrations are presented for a number of properties of the boson plasma as functions of temperature and density in the weak-coupling regime: these are the chemical potential, the condensate fraction, the normal and anomalous momentum distribution functions and the corresponding one-body density matrices, and the dispersion relation of single-particle excitations.
\end{abstract}
PACS numbers: 05.30.Jp, 03.75.Fi, 74.20.-z

\section{INTRODUCTION}
	The fluid of charged bosons over a uniform neutralizing background (CBF) interacting via inverse-first-power
repulsions is a model in quantum statistical mechanics\cite{1} having possible relevance to superconductivity in layered cuprate materials\cite{2} and to the equation of state and nuclear reactions in dense plasmas of pressure-ionized helium.\cite{3} Its counterpart with inverse-first-power attractions can arise in systems of atoms immersed in particular configurations of intense off-resonant laser beams.\cite{4}
	
	The properties of the CBF at zero temperature have been addressed in a number of theoretical studies, ranging from the early work on the weak-coupling regime\cite{5} to treatments based either on the use of Jastrow-Feenberg correlated wave functions\cite{6} or on self-consistent accounts of correlations.\cite{7} Quantum Monte Carlo calculations\cite{8} have covered the whole range of coupling strength up to Wigner crystallization. More recently, the evaluation of the condensate fraction has been extended to finite temperature\cite{9} within the theoretical frame provided by the Bogoliubov formalism. These theoretical results could only be tested against Monte Carlo data at zero temperature, showing that this most simple approach yields almost quantitative values for the interaction-induced depletion of the condensate over a non-negligible range of coupling strength.
	
	However, the standard Bogoliubov theory can be expected to fail with increasing temperature, as the noncondensate fraction increases. The purpose of the present work is to examine the roles of both the noncondensate and the anomalous Bose correlations in the CBF at finite temperature within the Hartee-Fock-Bogoliubov (HFB) approximation, with due attention to the chemical potential of the fluid as given by the Hugenholtz-Pines relation.\cite{10}
	
	It may be mentioned at this point that interest in the HFB approach to boson fluids at finite temperature has been revived in relation to developments in the study of Bose-Einstein condensed atomic gases under confinement. In this context Griffin\cite{11} has critically analyzed the HFB scheme with special regard to the classification of (conserving versus gapless) approximations given by Hohenberg and Martin\cite{12} for fluids of neutral bosons. The HFB approach has also been used to describe the shape deformation modes of confined atomic gases in dependence of temperature.\cite{13}
	
	In his work Griffin\cite{11} has emphasized that the problem of determining the excitation spectrum of a Bose-condensed fluid is difficult because of the need to satisfy several different constraints. In particular, in an explicit treatment of a Bose gas of neutral particles with contact interactions he has stressed that the HFB value for the chemical potential violates the exact Hugenholtz-Pines expression for this quantity and that the predicted excitation spectrum violates the gaplessness requirement. Proportionality between the single-particle and the collective sound-wave excitation spectra in the neutral fluid at long wavelengths can be understood from the argument developed by Gavoret and Nozi\`eres.\cite{14} The long-range nature of the Coulomb interactions in the CBF affects these properties and emphasizes the shortcomings of the HFB approach. Because of the plasmon gap in the collective excitation spectrum at long wavelengths, one expects the same gap in the single-particle spectrum, at least at $T=0$.\cite{5,15} As we shall see, the violation of the Hugenholtz-Pines relation is associated in the CBF to infrared divergencies and requires {\it ad hoc} mending. Analytical results can then be derived in the long-wavelength (low-momentum) region, extending exact results previously obtained\cite{15} from sum-rule arguments at zero temperature. The modified HFB approach, incorporating the Hugenholtz-Pines value of the chemical potential, is amenable to numerical evaluation and in particular displays the expected gap in the single-particle spectrum of the CBF at low temperature, though a discrepancy from the value of the plasmon gap remains.
	
	In brief, the plan of the paper is as follows. Section II collects for convenience the main standard equations of the HFB approach and presents their modification that we propose for a Bose plasma in order to incorporate the Hugenholtz-Pines relation. A further subsection presents our analytical results for the long-wavelength region and our numerical results for the limiting value of the excitation energy as a function of temperature in the same region. Various other numerical results are presented in Sec. III, while Sec. IV gives a brief summary and our conclusions.

\section{THE HFB APPROACH TO A BOSON PLASMA}

	As is well known,\cite{16} the ideal Bose gas starts undergoing Bose-Einstein condensation at a critical temperature $T_0=3.31(\hbar^2n^{2/3}/m k_B)$, where $n$ is the particle number density and $m$ the particle mass. The condensate density $n_0$ increases with decreasing temperature, to become equal to the total density $n$ in the ideal gas at zero temperature. The role of the interactions is to depress the condensate fraction and to shift the critical temperature $T_c$ away from $T_0$. Within the HFB approach this shift is a coupling-strength-dependent increase in the present case of Coulomb repulsions.
	
	The CBF is described by the Hamiltonian
\begin{equation}
H=\int d{\bf r}\psi^\dagger({\bf r})\left(-\frac{\hbar^2}{2m}\nabla^2-\mu \right)\psi({\bf r})+\frac{1}{2}\int\int d{\bf r}d{\bf r'}\psi^\dagger({\bf r'})\psi^\dagger({\bf r})V(|{\bf r}-{\bf r'}|)\psi({\bf r'})\psi({\bf r})
\end{equation}
where  $\psi({\bf r})$ is the bosonic field operator, $\mu$ is the chemical potential and $V(r)$ is the Coulomb interaction potential. The average potential felt by each particle vanishes because of the presence of the neutralizing background: that is, in the calculations reported below we take the Fourier transform of $V(r)$ as $V_k=4\pi e^2/k^2$ with $V_{k=0}$ set equal to zero\cite{151}. The coupling strength is measured by the dimensionless parameter $r_s$, defined by $r_s a_B=(4\pi n/3)^{-1/3}$ with $a_B$ the Bohr radius.

	The HFB approach is based on the equation of motion for the field operator and adopts a decoupling procedure in order to linearize it and solve it. The Bogoliubov prescription starts by separating out the condensate part from the field operator $\psi({\bf r})$ by setting $\psi({\bf r})=\psi_0\hat{a}_0+\hat{\phi}({\bf r})$, where $\hat{a}_0$ and $\hat{\phi}({\bf r})$ are the field operators for the condensate and for the cloud of particles promoted out of the condensate. Here, $\psi_0=\left\langle \psi({\bf r})\right\rangle$ is the condensate wave function, given by the ensemble average of $\psi({\bf r})$ in the symmetry-broken state. In the homogeneous gas we take the condensate wave function as uniform in space and time, and for the sake of simplicity we also assume that it is real.
	
	The equation of motion obeyed by the Heisenberg field operator is\cite{16}
\begin{equation}
i\hbar\frac{\partial \psi({\bf r},t)}{\partial t}=\left(-\frac{\hbar^2}{2m}\nabla^2-\mu \right)\psi({\bf r})+\int d{\bf r'}\psi^\dagger({\bf r'},t)\psi({\bf r'},t)\psi({\bf r},t)V(|{\bf r}-{\bf r'}|).
\end{equation}
Following a standard procedure, we use the Bogoliubov prescription in Eq. (2) to derive the equation of motion for the fluctuation operator $\hat{\phi}({\bf r})$.

	In the HFB scheme the following "mean field" approximation is adopted,\cite{11}
\begin{equation}
\left.
\begin{array}{l}
\hat{\phi}^\dagger({\bf r'},t)\hat{\phi}({\bf r},t)\cong\left\langle\hat{\phi}^\dagger({\bf r'},t)\hat{\phi}({\bf r},t) \right\rangle\equiv\tilde{n}(|{\bf r}-{\bf r'}|,t)\\ \\
\hat{\phi}({\bf r'},t)\hat{\phi}({\bf r},t)\cong\left\langle\hat{\phi}({\bf r'},t)\hat{\phi}({\bf r},t) \right\rangle\equiv\tilde{m}(|{\bf r}-{\bf r'}|,t)
\end{array}
\right\}
\end{equation}
where we have introduced the normal and the anomalous density matrix, $\tilde{n}$ and $\tilde{m}$. The three-point kernel $\hat{\phi}^\dagger({\bf r'},t)\hat{\phi}({\bf r'},t)\hat{\phi}({\bf r},t)$ is approximated as
\begin{equation}
\hat{\phi}^\dagger({\bf r'},t)\hat{\phi}({\bf r'},t)\hat{\phi}({\bf r},t)\cong\left\langle\hat{\phi}^\dagger({\bf r'},t)\hat{\phi}({\bf r'},t) \right\rangle\hat{\phi}({\bf r},t)+\tilde{n}(|{\bf r}-{\bf r'}|,t)\hat{\phi}({\bf r'},t)+\tilde{m}(|{\bf r}-{\bf r'}|,t)\hat{\phi}^\dagger({\bf r'},t)
\end{equation}
The equation of motion for the fluctuation operator in the HFB approximation then reads
\begin{equation}
i\hbar\frac{\partial \hat{\phi}({\bf r},t)}{\partial t}=\left(-\frac{\hbar^2}{2m}\nabla^2-\mu \right)\hat{\phi}({\bf r},t)+\int d{\bf r'}\left[n_0+\tilde{n}(|{\bf r}-{\bf r'}|,t)\right]\hat{\phi}({\bf r'},t)+\left[n_0+\tilde{m}(|{\bf r}-{\bf r'}|,t)\right]\hat{\phi}^\dagger({\bf r'},t)V(|{\bf r}-{\bf r'}|).
\end{equation}
By taking the average of Eq. (2) and using the approximations set out in Eqs. (3) and (4) we also obtain the expression for the chemical potential in the HFB approximation,
\begin{equation}
\mu_{HFB}=\int d{\bf r'}\left[\tilde{n}(|{\bf r}-{\bf r'}|,t)+\tilde{m}(|{\bf r}-{\bf r'}|,t)\right]V(|{\bf r}-{\bf r'}|).
\end{equation}
The condition of charge neutrality has been used in deriving Eqs. (5) and (6) by setting $V_{k=0}=0$, as already remarked.

The solution of Eq. (5) is achieved by means of a normal-mode expansion for the fluctuation operator,
\begin{equation}
\hat{\phi}({\bf r},t)=\sum_j\left[u_j({\bf r})\exp\left(-iE_jt/\hbar\right)\hat{a}_j+v_j^*({\bf r})\exp\left(iE_jt/\hbar\right)\hat{a}^\dagger_j\right].
\end{equation}
This leads to a set of Bogoliubov - de Gennes equations, which can be diagonalized in Fourier transform to obtain the expression for the HFB spectrum of single-particle excitations,
\begin{equation}
E_k=\sqrt{\left[\varepsilon_k+I_n(k)\right]^2-I_m(k)^2}
\end{equation}
and for the HFB mode amplitudes,
\begin{equation}
\left.
\begin{array}{l}
u_k^2=\frac{1}{2}\left[1+\left(\varepsilon_k+I_n(k)\right)/E_k\right]\\ \\
v_k^2=\frac{1}{2}\left[-1+\left(\varepsilon_k+I_n(k)\right)/E_k\right]
\end{array}
\right\}.
\end{equation}
In these equations $\varepsilon_k=\left(\hbar^2k^2/2m\right)-\mu$ and the self-energies $\left\{I_n(k),I_m(k)\right\}$ are to be determined self-consistently with Eqs. (8) and (9) according to
\begin{equation}
\left\{I_n(k),I_m(k)\right\}=n_0V_k+\sum_{{\bf q}\neq 0}V_{|{\bf q}-{\bf k}|}\left\{n(q),m(q)\right\}
\end{equation}
Here,
\begin{equation}
\left.
\begin{array}{l}
n(q)=v_q^2+f_q\left(u_q^2+v_q^2\right)\\ \\
m(q)=u_qv_q\left(2f_q+1\right)
\end{array}
\right\}
\end{equation}
with $f_q=\left[\exp{\left(E_q/k_BT\right)}-1\right]^{-1}$ are the Fourier transform of the normal and the anomalous density matrix. Evidently, $n(q)$ is the momentum distribution of the Bose fluid. Thus, the condensate fraction is given by
\begin{equation}
n_0=n-\sum_{{\bf q}\neq 0}\left[v_q^2+f_q\left(u_q^2+v_q^2\right)\right]
\end{equation}
Finally, the normal and anomalous one-body density matrices are given by
\begin{equation}
\left.
\begin{array}{l}
n(r)=n_0+\sum_{{\bf q}\neq 0}\left[v_q^2+f_q\left(u_q^2+v_q^2\right)\right]\exp\left(i{\bf q}\cdot {\bf r}\right)\\ \\
m(r)=n_0+\sum_{{\bf q}\neq 0}\left[u_qv_q\left(2f_q+1\right)\right]\exp\left(i{\bf q}\cdot {\bf r}\right)
\end{array}
\right\}.
\end{equation}

	Equations (8) - (10) reduce to the standard Bogoliubov approximation (see, {\it e.g.} Ref. [9]) when the contributions coming from the density matrices $n(q)$ and $m(q)$ are omitted. In the same approximation the chemical potential can be set to zero. In the HFB approximation, however, Eq. (6) for the chemical potential of the CBF is plagued by an infrared divergence (see below).
	
	We overcome this difficulty in the present case of a Bose plasma by recourse to the exact expression given for the chemical potential by the Hugenholtz-Pines relation. This states that
\begin{equation}
\mu=\Sigma_{11}({\bf q}=0,\omega=0)-\Sigma_{12}({\bf q}=0,\omega=0)
\end{equation}
where $\Sigma_{ij}({\bf q},\omega)$ is the $2\times 2$ matrix for the self-energies of a Bose-condensed gas.\cite{12} We then replace the divergent expression in Eq. (6) by the expression
\begin{eqnarray}\label{muHP}
\mu_{HP}& &=\int d{\bf r'}\left[\tilde{n}(|{\bf r}-{\bf r'}|,t)-\tilde{m}(|{\bf r}-{\bf r'}|,t)\right]V(|{\bf r}-{\bf r'}|)\nonumber\\& &=\sum_{{\bf q}\neq 0}V_q\left[n(q)-m(q)\right].
\end{eqnarray}
Equation (15) follows from Eq. (14) by using the HFB expression for the self-energies given in Eq. (10), together with the requirement of charge neutrality which suppresses the ${\bf q}=0$ term in the sum.

	Equations (8) - (11) are well behaved when one adopts the expression in Eq. (15) for the chemical potential. Their solution will be presented in Sec. III below. However, before proceeding to numerical calculations we need to evaluate analytically the long-wavelength behavior of the CBF in the HFB approach.
\subsection{Long-wavelength behaviors}

	The following asymptotic expressions will be demonstrated below to hold at low momenta for the CBF at temperature $T\geq 0$ in the HFB approximation:
\begin{equation}
\lim_{k\rightarrow 0}\left[I_n(k)=I_m(k)\right]=\frac{4\pi e^2n_0}{k^2}+\frac{\left(4\pi e^2\right)^2n_0}{16E_0k}(2f_0+1)
\end{equation}
where we have introduced the notations $E_0=E_{k=0}$ and $f_0=f_{k=0}$. The infrared divergence in the HFB expression for the chemical potential in Eq. (6), and the cancellation of the divergence at both leading and subleading order in Eq. (15) from the Hugenholtz-Pines relation, are immediately evident.

	The corresponding asymptotic expressions for the normal and anomalous distribution are obtained from Eqs. (9) - (11), which yield
\begin{equation}
\lim_{k\rightarrow 0}\left[n(k)=m(k)\right]=(2f_0+1)\lim_{k\rightarrow 0}\left[I_n(k)/2E_k\right].
\end{equation}	
The result is
\begin{equation}
\lim_{k\rightarrow 0}\left[n(k)=m(k)\right]=(2f_0+1)\left[\frac{4\pi e^2n_0}{2E_0k^2}+\frac{\left(4\pi e^2\right)^2n_0}{32E_0^2k}(2f_0+1)\right].
\end{equation}
This asymptotic expression for $n(k)$ reduces at $T=0$ to the exact result obtained by sum rule arguments in Ref. [15], provided that the energy $E_0$ coincides with the plasma frequency $\omega_p=\sqrt{4\pi ne^2/m}$. We present at the end of this section a calculation of $E_0$ in the HFB approach and compare the result with the plasma frequency.

	To demonstrate the results given above in Eqs. (16) and (18), we examine the HFB expressions for the self-energies $\left\{I_n(k),I_m(k)\right\}$ in Eq. (10). The integral on the RHS of this equation is easily shown not to contribute at order $k^{-2}$, so that the leading-order term in Eq. (16) is immediately established. The leading-order term in Eq. (18) follows from Eq. (17).
	
	With the leading-order expression for $n(k)$ and $m(k)$ we can now return to the evaluation of the integral on the RHS of Eq. (10) and determine the subleading term in Eq. (16). For the quantity $I_n(k)$ we write
\begin{eqnarray}
\sum_{{\bf q}\neq 0}V_{|{\bf k}-{\bf q}|}n(q)=& &4\pi e^2\sum_{{\bf q}\neq 0}\frac{1}{k^2-2{\bf k}\cdot{\bf q}+q^2}\left[n(q)-\lim_{q\rightarrow 0}n(q)\right]\nonumber\\ & &+\frac{\left(4\pi e^2\right)^2n_0}{2E_0}(2f_0+1)\sum_{{\bf q}\neq 0}\frac{1}{q^2\left(k^2-2{\bf k}\cdot{\bf q}+q^2\right)}
\end{eqnarray}
The value of the second integral on the RHS of Eq. (19) is $1/(8k)$, while the first integral does not contribute to this order.\cite{17} With an identical argument for $I_m(k)$ the subleading-order terms in Eq. (16), and hence in Eq. (18), are thus established.

	We turn next to the evaluation of the energy $E_0$ from Eq. (8) in the long-wavelength limit. We get
\begin{equation}
E_0/\hbar=\left(\frac{4\pi e^2n_0}{m}\right)^{1/2}\lim_{k\rightarrow 0}\left\{1+\frac{2m}{\hbar^2k^2}\left[I_n(k)-I_m(k)-\mu\right]\right\}^{1/2}.
\end{equation}
We also have from Eqs. (10) and (14) that
\begin{equation}
\lim_{k\rightarrow 0}\left[I_n(k)-I_m(k)-\mu\right]=\frac{1}{3}k^2\sum_{{\bf q}\neq 0}q^{-2}V_q\left[n(q)-m(q)\right]
\end{equation}
where according to Eq. (11) we have
\begin{equation}
n(q)-m(q)=-\frac{1}{2}+\frac{\hbar^2q^2}{4mE_q}(2f_q+1)
\end{equation}
so that
\begin{equation}
\lim_{q\rightarrow 0}\left[n(q)-m(q)\right]=-\frac{1}{2}.
\end{equation}
We can thus write the final expression for $E_0$ in a form which is suitable for numerical calculation,
\begin{equation}
E_0/\hbar=\left(\frac{4\pi e^2n_0}{m}\right)^{1/2}\left\{1+\frac{2m}{3\hbar^2}\sum_{q\neq 0}q^{-2}V_q\left[n(q)-m(q)+\frac{1}{2}\right]\right\}^{1/2}.
\end{equation}
This expression shows that the value of $E_0$ is finite, as implicitly assumed in the foregoing derivation of Eqs. (16) and (18). The standard Bogoliubov approximation yields $E_0/\hbar=(4\pi e^2n_0/m)^{1/2}$. As is shown in the left panel in Figure 1, the Hartree-Fock terms inside the brackets in Eq. (24) raise the value of $E_0/\hbar$ towards the plasma frequency $(4\pi e^2n/m)^{1/2}$ at low temperature, without quite reaching it. In both the Bogoliubov and the HFB approximation the value of $E_0/\hbar$ drops with increasing temperature to vanish at the critical temperature $T_c$. The corresponding behaviour of the chemical potential from Eq. (\ref{muHP}) is shown in the right panel in Figure 1.

	It is also worth making a remark on the asymptotic result given in Eq. (23). The quantity $n(q)-m(q)$ is the HFB value for a special component of the two-body density matrix in the CBF, that is $n(q,q)/n_0$. It was shown in Ref. [15] from an exact sum-rule argument that this quantity takes the value $-1/2$ at low momenta in the CBF at $T=0$. Thus, this exact asymptotic result holds in the HFB approximation, which yields the same limiting value at all temperatures $T<T_c$.

\section{NUMERICAL RESULTS}

	We present in this section the results of numerical calculations of some other properties of the weakly coupled CBF in the modified HFB approximation. Figure 2 reports the condensate fraction $n_0/n$ as a function of temperature at coupling strength $r_s=1$. In comparison with the result of the simple Bogoliubov (B) approach, there is a further large shift in the value of the critical temperature for the formation of a condensate away from the ideal-gas value. At $T=0$ the change in the interaction-induced depletion of the condensate from the result of the simple Bogoliubov (B) approach is quite small at low $r_s(r_s\leq 5, \ {\rm say})$, thus confirming the results reported in Ref. [9] and ensuring good agreement with the Monte Carlo data obtained by Moroni {\it et al}.\cite{8} However, with further increase of $r_s$ the HFB condensate fraction drops very rapidly, in disagreement with the Monte Carlo data.
	
	In view of the asymptotic behaviors given in Eq. (18), Figure 3 reports the quantity $q^2n(q)$ at $T=0$ in order to display the subleading term which is absent in the simple Bogoliubov approximation. This term is also evident in the Monte Carlo data, which are reported in Figure 3 from the work of Moroni {\it et al}.\cite{8}, and there is in fact reasonable quantitative agreement between the HFB momentum distribution and these data.The HFB predictions for the normal one-body density matrix $n(r)$ at various temperatures are shown in Figure 4.
	
	Similar results are reported for the anomalous Bose correlations in Figures 5 and 6. Figure 5 shows both $q^2n(q)$ and $q^2m(q)$ at $r_s=1$ and two temperatures, to illustrate how departures arise with increasing momentum away from their common limiting values in Eq. (18). The normal and anomalous one-body density matrices are given in Figure~6.
	
	Finally, Figure 7 reports the HFB dispersion relation of single-particle excitations as a function of temperature at $r_s=1$. The deviations from the plasma frequency in the long-wavelength limit and their dependence on temperature have already been displayed in Figure 1. The Figure shows that, aside for this, there is very little dependence of the dispersion relation on temperature. There is, of course, no damping of these excitations within the HFB approach.

\section{CONCLUSIONS}

	Summarizing, we have examined the results given by the Hartree-Fock-Bogoliubov approximation for the quasiparticle excitations in a weakly coupled boson plasma through the temperature range in which a Bose condensate is present. The long-range nature of the Coulomb interactions emphasizes the difficulties of the theory that in previous work on neutral boson fluids have been shown to arise from subtle dynamical correlations induced by the presence of a condensate. For the boson plasma the Coulomb interactions generate infrared divergencies in both the normal and the anomalous momentum distribution, which are reproduced in the HFB approach at both leading and subleading order. However, the Coulomb case requires an {\it ad hoc} adjustment of the chemical potential. Exact cancellation of divergent terms occurs between the two distributions in the evaluation of the quasiparticle excitation energy at long wavelengths. The residual non-divergent HFB terms raise the quasiparticle energy towards the plasma frequency at low temperature, without quite satisfying the basic requirement that the Coulomb-induced gap in the quasiparticle spectrum at long wavelengths should coincide with the plasmon gap in the collective density-fluctuation spectrum.
	
	It appears, therefore, that the HFB approach may not provide an entirely fruitful starting point for an internally consistent description of the boson plasma. An approach based on the so-called dielectric formalism\cite{18} may be more promising and we hope to examine it in future work, with specific attention to the relationship between collective and single-particle excitation frequencies as a function of temperature.
\acknowledgements
{
This work was partially supported by MIUR under the PRIN-2000 Initiative.
}

\newpage 

\begin{figure}
\centerline{\mbox{\psfig{figure=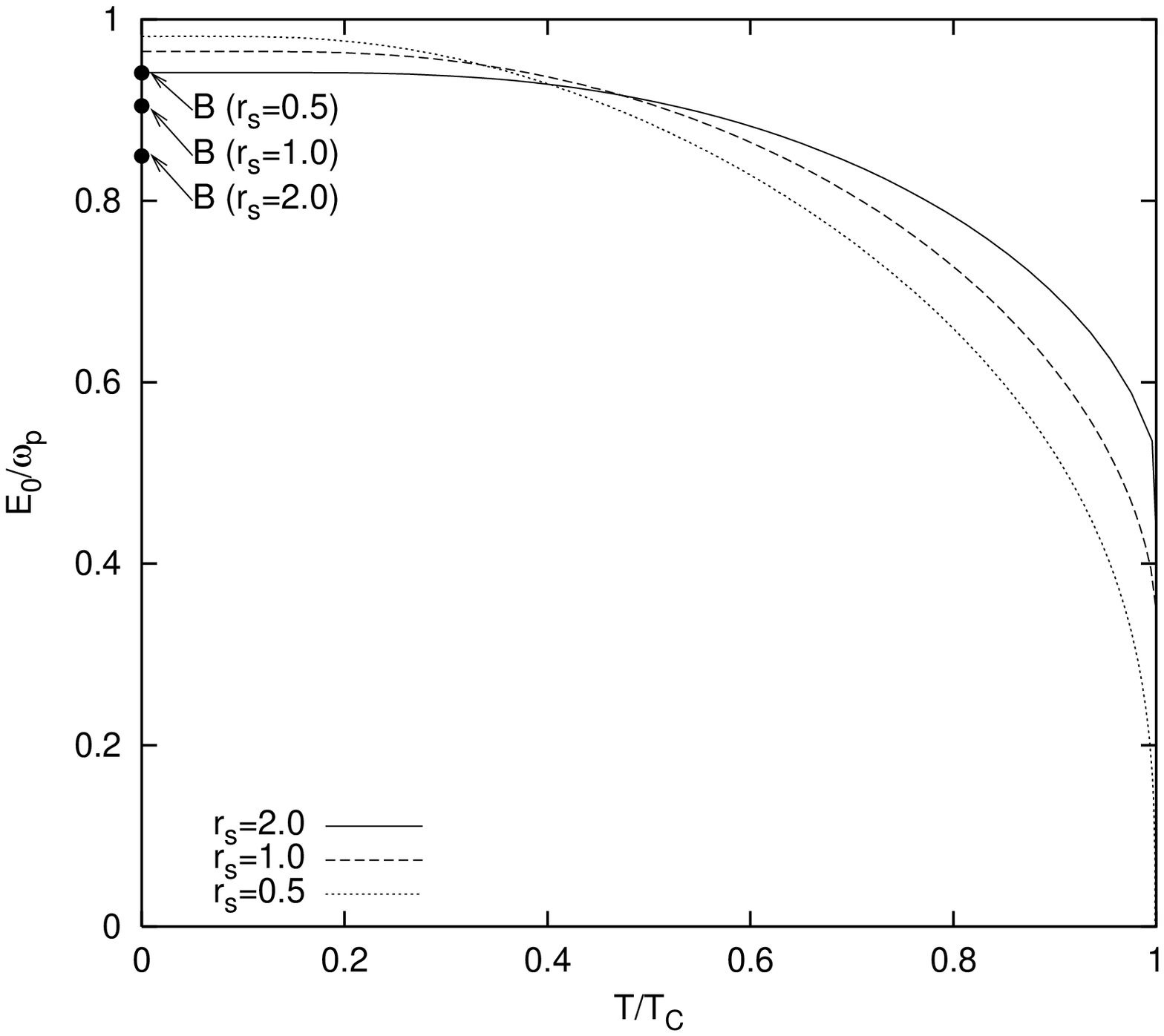, angle =0, width =9 cm}\psfig{figure=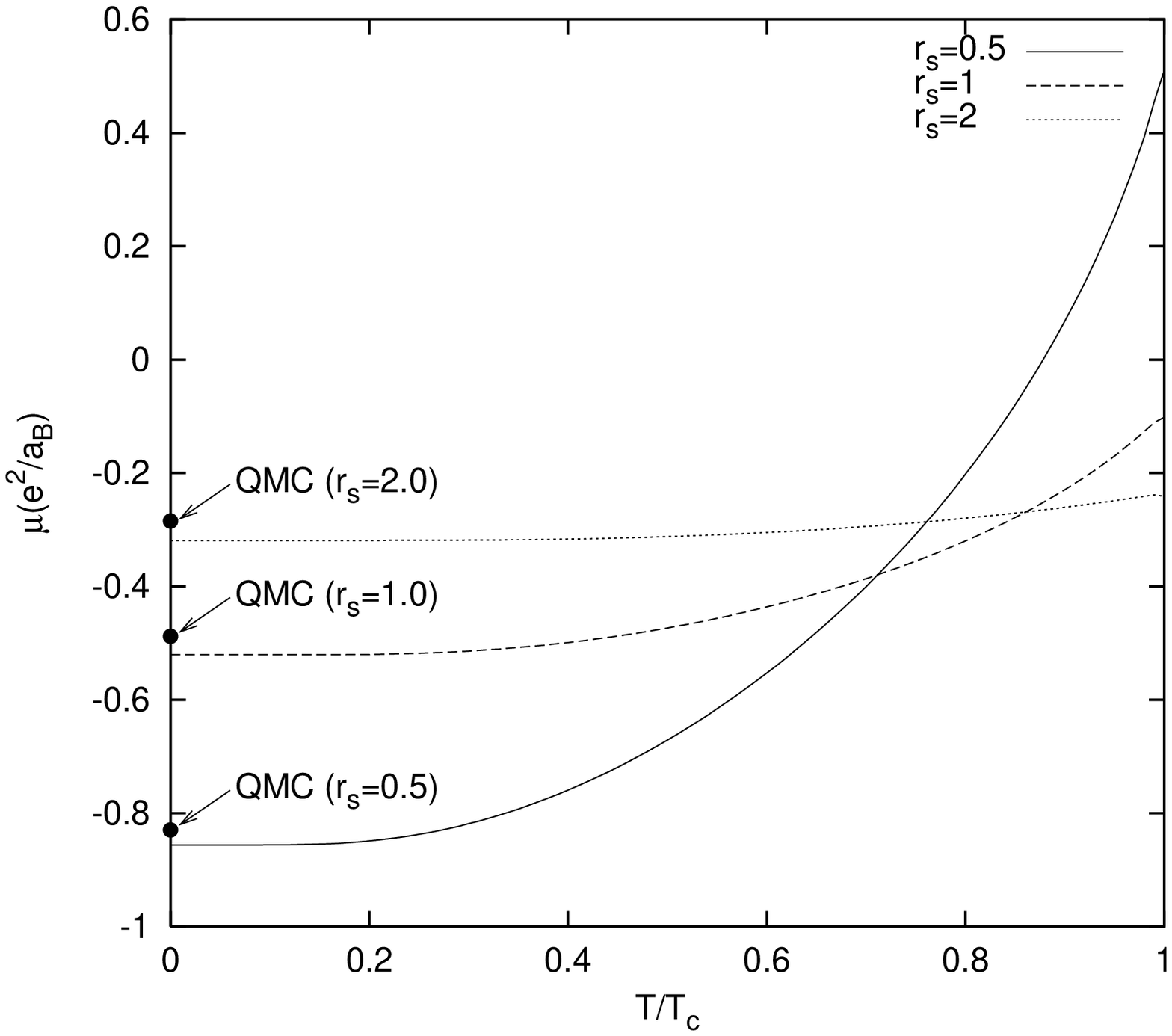, angle =0, width =9 cm}}} 
\caption{Left: the HFB single-particle excitation frequency $E_0$ at long wavelengths (in units of the plasma frequency $\omega_p$) as a function of temperature $T/T_c$ (in all figures $T_c$ is the HFB critical temperature) for three values of the coupling strength $r_s$. The dots marked by B show the corresponding values in the standard Bogoliubov approximation at $T=0$. Right: curves for the chemical potential from the Hugenholtz-Pines relation. The QMC values reported at $T=0$ are from Moroni {\it et al}.\cite{8}}
\label{Fig1}
\end{figure}

\begin{figure}
\centerline{\mbox{\psfig{figure=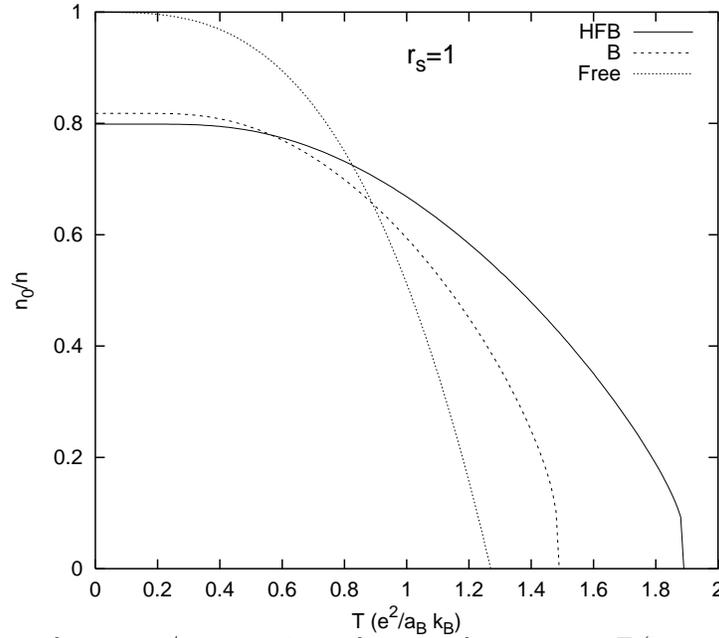, angle =0, width =10 cm}}} 
\caption{The HFB condensate fraction $n_0/n$ at $r_s=1$ as a function of temperature $T$ (in units of $e^2/a_Bk_B)$), compared with the results for the ideal boson gas (Free) and for the Bogoliubov approach (B).}
\label{Fig2}
\end{figure}

\begin{figure}
\centerline{\mbox{\psfig{figure=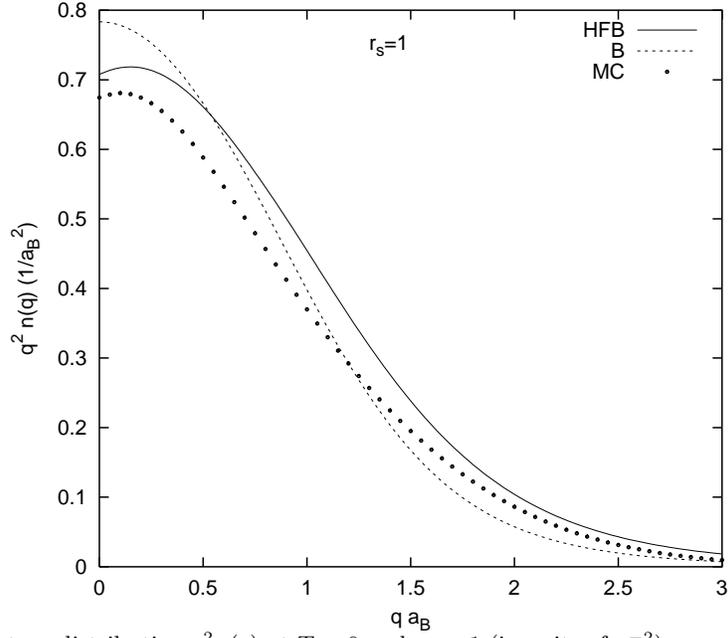, angle =0, width =10 cm}}} 
\caption{The HFB momentum distribution $q^2n(q)$ at $T=0$ and $r_s=1$ (in units of $a_B^{-2}$), compared with the results from the Bogoliubov approach (B) and from diffusion Monte Carlo runs (MC).}
\label{Fig3}
\end{figure}

\begin{figure}
\centerline{\mbox{\psfig{figure=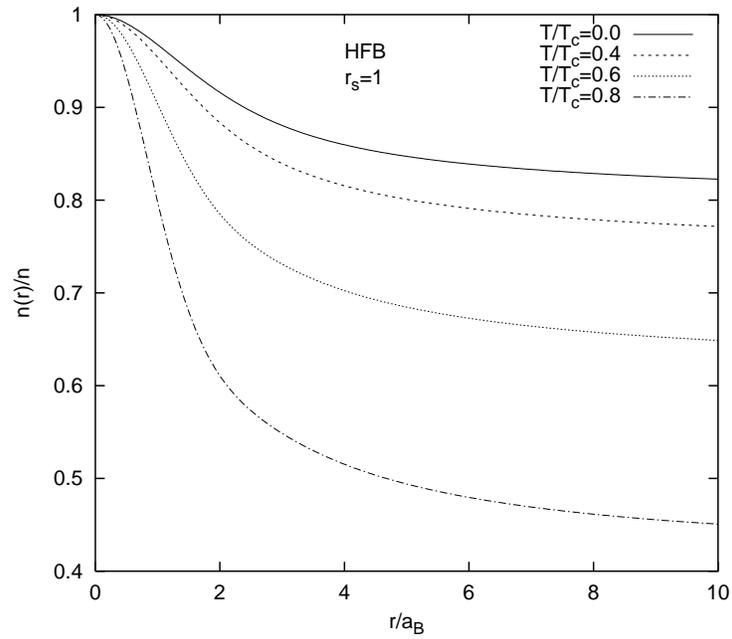, angle =0, width =10 cm}}} 
\caption{The HFB normal density matrix $n(r)/n$ at $r_s=1$ and various values of the reduced temperature $T/T_c$.}
\label{Fig4}
\end{figure}

\begin{figure}
\centerline{\mbox{\psfig{figure=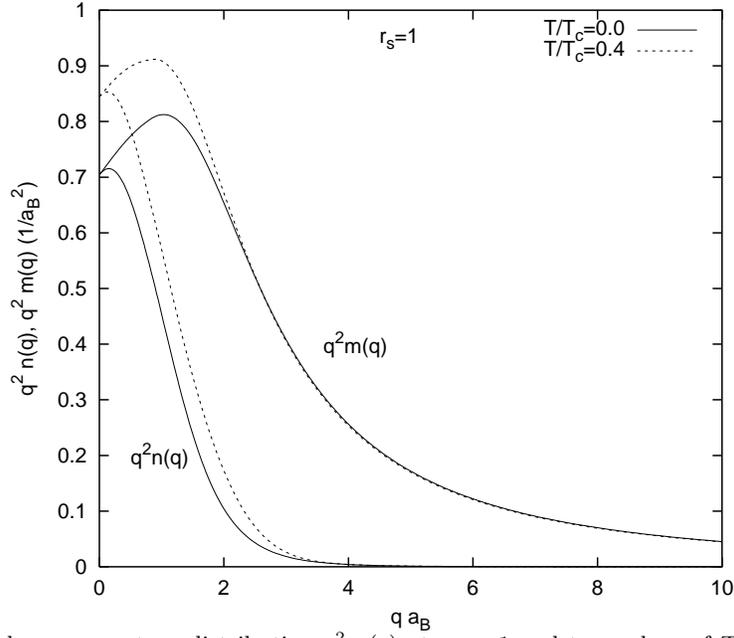, angle =0, width =10 cm}}} 
\caption{The HFB anomalous momentum distribution $q^2m(q)$ at $r_s=1$ and two values of $T/T_c$, compared with the HFB normal momentum distribution $q^2n(q)$. Both curves are in units of $a_B^{-2}$.}
\label{Fig5}
\end{figure}

\begin{figure}
\centerline{\mbox{\psfig{figure=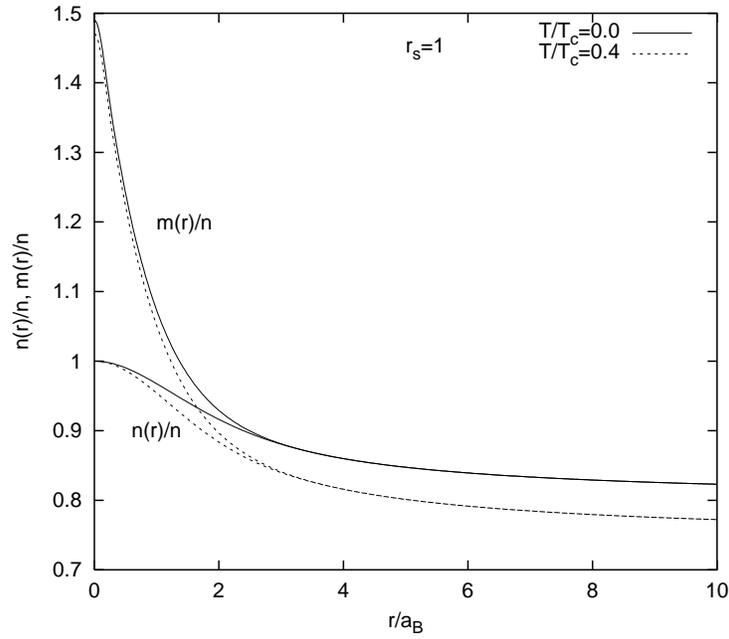, angle =0, width =10 cm}}} 
\caption{The HFB normal and anomalous density matrices at $r_s=1$ and two values of $T/T_c$.}
\label{Fig6}
\end{figure}

\begin{figure}
\centerline{\mbox{\psfig{figure=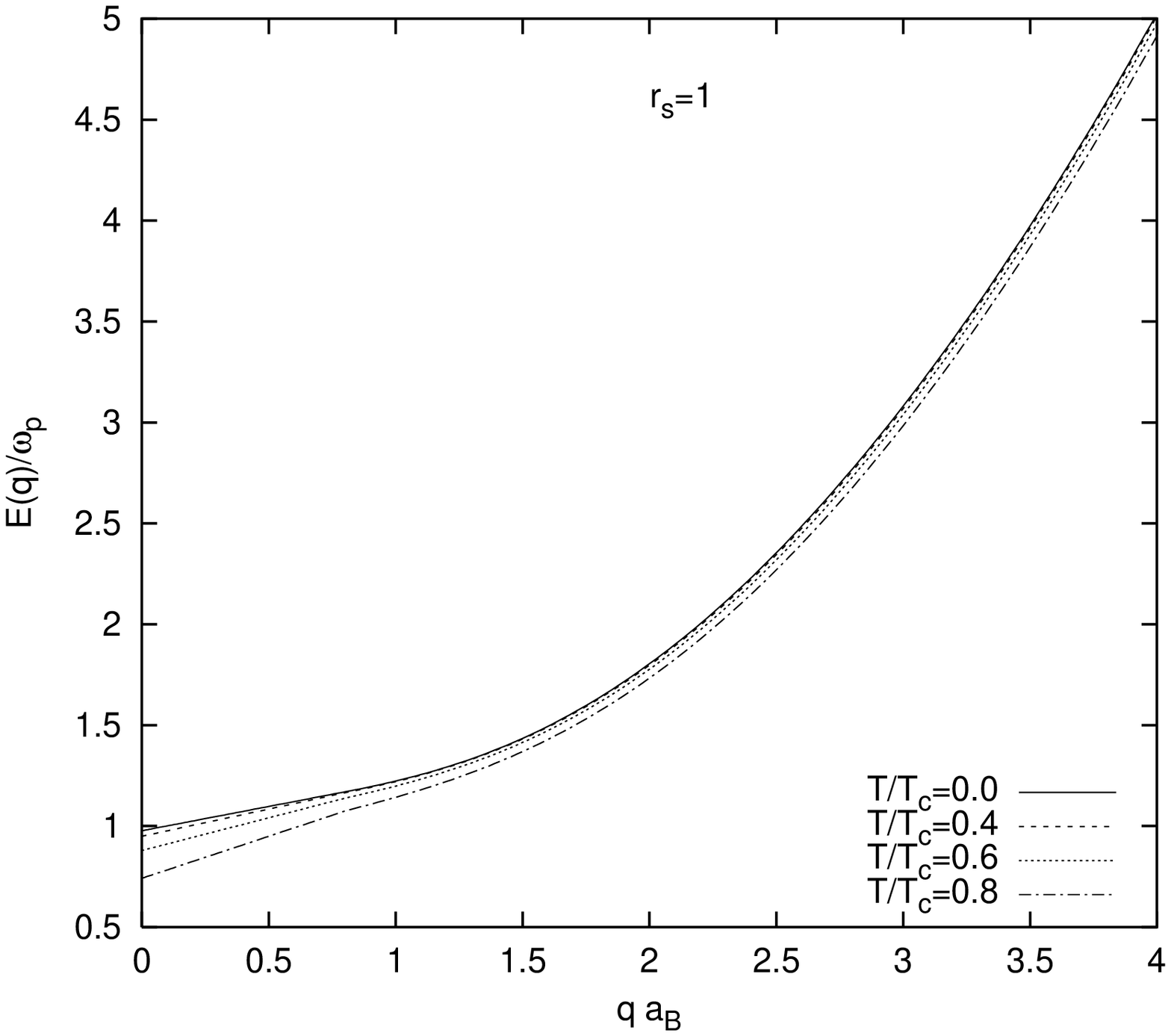, angle =0, width =10 cm}}} 
\caption{The HFB dispersion relation for single-particle excitations (in units of the plasma frequency $\omega_p$) at $r_s=1$ and various values of $T/T_c$.} 
\label{Fig7}
\end{figure}

\begin{references}
\bibitem{1}
F.Y. Wu and E. Feenberg, Phys. Rev. {\bf 128}, 943 (1962).
\bibitem{2}  	
M.R. Schafroth, Phys. Rev. {\bf 96}, 1149 (1954); A.S. Alexandrov and N.F. Mott, Supercond. 	Sci. Techn. {\bf 6}, 215 (1993); A.S. Alexandrov and P.P. Edwards, Physica C {\bf 331}, 97 (2000).
\bibitem{3}
B.W. Ninham, Phys. Lett. {\bf 4}, 278 (1963); V.L. Ginzburg, J. Stat. Phys. {\bf 1}, 3 (1969); J.P. 	Hansen, B. Jancovici, and D. Schiff, Phys. Rev. {\bf 29}, 991 (1972); S. Schramm, K. Langange, 	and S.E. Koonin, Astrophys. J. {\bf 397}, 579 (1992); H.-M. M\"uller and K. Langange, Phys. Rev. 	C {\bf 49}, 524 (1996).
\bibitem{4}  
D. O'Dell, S. Giovanazzi, G. Kurizki, and V.M. Akulin, Phys. Rev. Lett. {\bf 84}, 5687 (2000).
\bibitem{5}  
L.L. Foldy, Phys. Rev. {\bf 124}, 649 (1961); K.A. Br\"uckner, Phys. Rev. {\bf 156}, 204 (1967); S.K. 	Ma and C.W. Woo, Phys. Rev. {\bf 159}, 165 (1967).
\bibitem{6}  
D.K. Lee, Phys. Rev. {\bf 187}, 326 (1969); D.K. Lee and F.H. Ree, Phys. Rev. A {\bf 5}, 814 (1972); 	R. Monnier, Phys. Rev. A {\bf 6}, 393 (1972); J.P. Hansen and R. Mazighi, Phys. Rev. A {\bf 18}, 	1282 (1978); M. Saarela, Phys. Rev. B {\bf 29}, 191 (1984); V. Apaja, J. Halinen, V. Halonen, E. 	Krotscheck, and M. Saarela, Phys. Rev. B {\bf 55}, 12925 (1997).
\bibitem{7}  
A. A. Caparica and O. Hip\'olito, Phys. Rev. A {\bf 26}, 2832 (1982); A. Gold, Z. Phys. B {\bf 89}, 1 	(1992); S. Conti, M.L. Chiofalo, and M.P. Tosi, J. Phys.: Condens. Matter {\bf 6}, 8795 (1994); 	A. Yu. Cherny and A.A. Shanenko, Phys. Lett. A {\bf 250}, 170 (1998).
\bibitem{8}  
D.M. Ceperley and B.J. Alder, Phys. Rev. Lett. {\bf 45}, 566 (1980); D.M. Ceperley and B.J. 	Alder, J. Phys. (Paris), Colloq. {\bf 7}, C295 (1980); G. Sugiyama, C. Bowen, and B.J. Alder, 	Phys. Rev. B {\bf 46}, 13042 (1992); S. Moroni, S. Conti, and M.P. Tosi, Phys. Rev. B {\bf 53}, 9688 	(1996).
\bibitem{9}  
E. Strepparola, A. Minguzzi, and M.P. Tosi, Phys. Rev. B {\bf 63}, 104509 (2001).
\bibitem{10}  
N. Hugenholtz and D. Pines, Phys. Rev. {\bf 116}, 489 (1959).
\bibitem{11}
A. Griffin, Phys. Rev. B {\bf 53} 9341 (1996).
\bibitem{12}
P.C. Hohenberg and P.C. Martin, Ann. Phys. (N.Y.) {\bf 34} 291 (1965).
\bibitem{13}
D.A.W. Hutchinson, E. Zaremba, and A. Griffin, Phys. Rev. Lett. {\bf 78} 1842 (1997); R.J. Dodd, 	M.
Edwards, C.W. Clark, and K. Burnett, Phys. Rev. A {\bf 57}, R32 (1998); H. Shi and W.-M. 	Zheng, Phys. Rev. A {\bf 59}, 1562 (1999).
\bibitem{14}
J. Gavoret and P. Nozi\`eres, Ann. Phys. (NY) {\bf 28}, 349 (1964).
\bibitem{15}
M. L. Chiofalo, S. Conti, and M. P. Tosi, J. Phys.: Condens. Matter {\bf 8}, 1921 (1996).
\bibitem{151}
E. H. Lieb and H. Narnhofer, J. Stat. Phys. {\bf 12}, 291 (1975); M. Rovere, G. Senatore, and M. P. Tosi in {\it Progress
on Electron Properties of Solids}, ed. R. Girlanda (Kluwer, Dordrecht, 1989) p. 221.
\bibitem{16}
See, {\it e.g.} A.L. Fetter and J.D. Walecka, {\it Quantum Theory of Many Particle Systems} 	(McGraw-Hill, New York, 1971).
\bibitem{17}
The form of the sub-subleading term in Eq. (16) can be assessed from the first integral on 	the RHS of Eq. (19) to be $\ln(k)$. To show this, we use the form of the subleading term in $n(q)$, which is of order $q^{-1}$, and after performing the angular integration we break the integral over $|{\bf q}|$ into the sum of an integral from $0$ to $k$ and an integral from $k$ to $\infty$.
\bibitem{18}
See, {\it e.g.} A. Griffin, {\it Excitations in a Bose-Condensed Liquid} (Cambridge University Press, 	New York, 1993), Sec. 5.1.
\end{references}
\end{document}